\documentclass[prb,aps,twocolumn,showpacs]{revtex4}
\usepackage{graphicx}

\begin{document}
\title{Coherent spin dynamics of an interwell excitons gas
in GaAs/AlGaAs coupled quantum wells}
\author{A.~V.~Larionov and V.~E.~Bisti}
\email{larionov@issp.ac.ru}
\affiliation{Institute of Solid State
Physics, Russian Academy of Sciences, 142432, Chernogolovka, Russia}
\author{M.~Bayer}
\affiliation{Experimentelle Physik II, Universit\"at Dortmund,
D-44221 Dortmund, Germany}
\author{J.~Hvam and K.~Soerensen}
\affiliation{Microelectronic Centret, Technical University of
Denmark, DK 2800 Lyngby, Denmark}

\date{\today}
\begin{abstract}
The spin dynamics of an interwell excitons gas has been
investigated in n-i-n GaAs/AlGaAs coupled quantum wells (CQWs). In
these heterostructures the electron and the hole are spatially
separated in neighboring quantum wells by a narrow AlAs barrier,
when an electric field is applied. The time evolution kinetics of
the interwell exciton photoluminescence has been measured under
resonant excitation of the 1sHH intrawell exciton, using a pulsed
tunable laser. The formation of a collective exciton phase in time
and the temperature dependence of its spin relaxation rate have
been studied. The spin relaxation rate of the interwell excitons
is strongly reduced in the collective phase. This observation
provides evidence for the coherence of the indirect excitons
collective phase at temperatures below a critical $T_c$.
\end{abstract}

\maketitle

\section{Introduction}

Among the quasi-two-dimensional systems based on semiconductor
heterostructures, coupled quantum wells are of special interest
because they may provide a spatial separation of photoexcited
electrons and holes in neighboring quantum wells [1]. For example,
in n-i-n type GaAs/AlGaAs CQWs with tilted bands due to bias
application, excitons can be excited with electron and hole
confined in adjacent wells which are separated by a tunneling
barrier. These excitons are called spatially indirect or interwell
excitons (IEs) and differ from the direct intrawell excitons
(DEs), for which electron and hole are located in the same QW. In
contrast to intrawell excitons, IEs are long-lived because the
wave functions of electron and hole overlap very weakly through
the tunneling barrier. This might open the possibility for such an
electron-hole system to maintain electron spin orientation as long
as the IEs life-time (several nanoseconds and longer). A large
number of IEs can be easily accumulated and this exciton gas can
be cooled down to rather low temperatures. Various possible
scenarios of collective behavior of a dense system of spatially
separated electrons and holes have been considered theoretically
[1-3]. Further, there are already a lot of works [see reviews 4-6]
reporting on collective behavior of IEs upon reaching critical
conditions.

Earlier, we have found that below a critical temperature the gas
of IEs in CQWs undergoes a phase transition-like behavior with
increasing exciton density  [7]. Experimental findings such as
strong narrowing of the IEs photoluminescence line, drastic
increasing of its circular polarisation degree and high
sensitivity with respect to temperature have been associated with
the condensation of IEs to a collective dielectric phase. Later it
has been shown that if critical conditions are satisfied the IEs
collective phase is most likely to occur in domain regions with
lateral confinement [8]. According to our experiments the
condensation occurs at $T < 4$K for an average exciton
concentration of $n_{ex}\sim 3 \times 10^{10}$cm$^{-2}$.

A collective excitonic phase, corresponding to a macroscopic
exciton occupation of the lower state in domain, should show
spatial and temporal coherence. This means that within the
coherence length condensed excitons are described by a common wave
function. Consequences expected from this are an increase of the
radiative decay rate of the excitons and a reduction of the
exciton spin relaxation rate. Due to these features the
opportunity for resonant photoexcitation of a spin aligned
collective interwell excitonic phase might arise. In the
considered case the coherence length scale is expected to be equal
to the size of the domain, arising from long range potential
fluctuations (around one micron in lateral size), in which IEs can
be accumulated.

In this manuscript we will address the IEs spin relaxation rate by
measuring and analyzing the circular polarization degree under
resonant pulsed laser photoexcitation. The paper is organized as
follows. After describing the CQWs heterostructures being studied
as well as the experimental technique in Section 2A, we describe
in Section 2B the time evolution of the spectra and the decay
curves of the luminescence intensity of IEs under conditions of
resonant pulsed laser excitation. A broad set of experimental
parameters (pump power, electrical bias, and temperature) was
varied. In Section 3 we discuss theoretical calculations of the
IEs spin relaxation rate. Finally in the concluding section 4
various properties of IEs observed in their luminescence spectra
and their critical behavior as function of optical pumping and
temperature are interpreted in terms of a collective exciton
behavior.

\begin{figure}[h]
\begin{center}
\includegraphics[width=8.5cm]{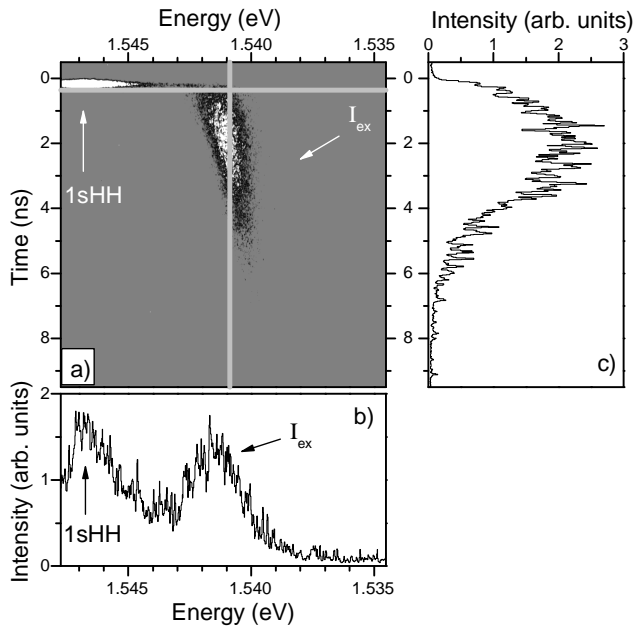}
\end{center}
\caption{(a) Streak camera image of the photoluminescence signal
as result of excitation of the sample by a short laser pulse
(about 1 ps duration). The brightness corresponds to the
photoluminescence intensity. The horizontal axis gives energy, the
vertical axis gives time. (b) Horizontal cut through the image (as
indicated by the corresponding line in panel (a)) gives the
photoluminescence spectrum at a fixed time. (c) Vertical cut (as
for the vertical line in (a)) gives the photoluminescence kinetics
at a fixed wavelength. Arrows indicate signal from the intrawell
exciton (1sHH) and the interwell exciton ($I_{IE}$). The image was
obtained for a bias voltage U = 0.65V at T = 1.85K.}
\end{figure}

\section{Experiment}

\subsection{Samples and experimental setup}

We have investigated a n-i-n GaAs/AlGaAs heterostructure
containing a GaAs/AlAs/GaAs CQW with a width of the GaAs wells of
about $120\AA$, and a width of the AlAs barrier of about $11\AA$.
The structure was grown using molecular-beam epitaxy on a $n$-type
doped GaAs substrate (concentration of the Si-impurity doping
$\sim 10^{18}$cm$^{-3}$) with (001) crystallographic orientation.
First, a 0.5$\mu$m thick buffer layer of Si-doped
($10^{18}$cm$^{-3}$) GaAs was grown on top of the substrate. Next,
an insulating AlGaAs layer ($x = 0.33$) with a thickness of
0.15$\mu$ was deposited. Then the GaAs/AlAs/GaAs CQWs sequence was
grown. To improve the interface quality, the growth interruption
technique has been used for the AlAs heteroboundaries. After the
CQW, an insulating AlGaAs layer with a thickness of 0.15$\mu$ was
grown, followed by a 0.1$\mu$ thick layer of Si-doped
($10^{18}$cm$^{-3}$) GaAs. The whole structure was covered by a
$100\AA$ wide GaAs layer. Mesas with a lateral size $1 \times 1
$mm$^{2}$ were prepared on this structure by lithography. Further,
Au + Ge + Pt alloy metallic contacts to the buffer layer and to
the doped layer in the upper part of a mesa were evaporated.

\begin{figure}[h]
\begin{center}
\includegraphics[width=8.5cm]{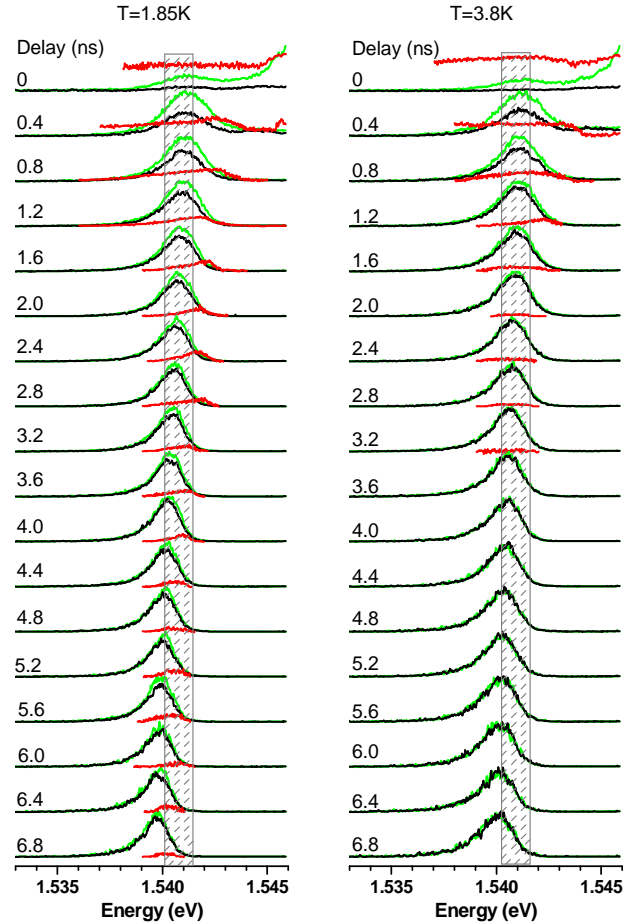}
\end{center}
\caption{Time-resolved IEs PL spectra taken at T = 1.85K and  T =
3.8K for various delay times after laser pulse excitation.
$U$=0.6V.}
\end{figure}

The IE photoluminescence (PL) was excited by 120-femtosecond laser
pulses with a repetition rate of 76 MHz. A holographic grating
with optical slits has been used for pulse shaping. The detection
of the signal was provided by a Hamamatsu streak-camera (Model
5680-24) with a Si charged-coupled-device (CCD) detector attached
to a 0.5-m spectrometer (Acton SP-500i). The systems time
resolution was about 70 ps in this configuration. For circular
polarization analysis of the PL signal under resonant
photoexcitation we have used linear polarizers and quarter wave
retarder plates.

\begin{figure}[h]
\begin{center}
\includegraphics[width=6.5cm]{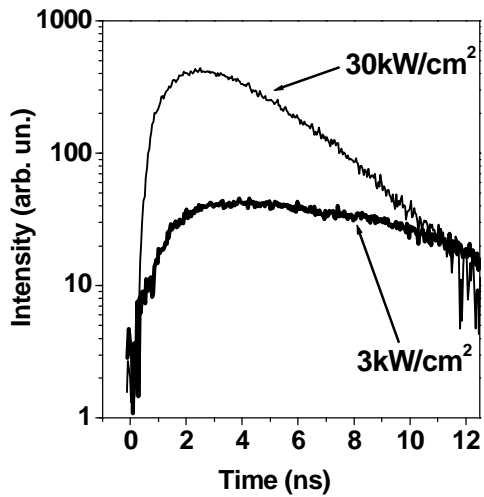}
\end{center}
\caption{IEs PL decay curves measured under pulsed photoexcitation
at different excitation powers, detected at the PL line maximum
(T=2K, U=0.6V).}
\end{figure}

\subsection{Experimental results.}

Fig. 1(a) gives a contour plot of a streak camera image of the PL
emission from the studied CQWs. The horizontal (vertical) axis
gives energy (time), while the brightness gives the PL intensity.
For data analysis this image has been profiled either along the
energy axis (fig.1b) or along the time-axis (fig.1c), resulting in
time-resolved PL spectra and energy-resolved PL decay curves,
respectively. For spin orientation of the IEs we used circular
polarized (for example, $\sigma^{+}$) laser excitation resonant to
the ground state of the intrawell 1sHH excitons. The interwell
exciton PL kinetics was measured under these conditions for
different temperatures and bias voltages.

Figure 2 shows the time evolution of the polarization resolved PL
spectra ($\sigma^{+}$ - green curves and $\sigma^{-}$ - black
curves) as well as the circular polarization degrees across the PL
emission spectra (red curves). The spectra have been measured for
various time delays relative to the exciting laser pulse at
$T$=1.8 and 3.8K for an applied voltage $U$=0.6V. At zero delay,
the IE PL is strongly circularly polarized, following the
polarization of the exciting laser, and has a full width at half
maximum (FWHM) of about 3meV. As the delay increases the PL line
narrows and shifts somewhat toward the long-wavelength part of the
spectrum. At $T$=1.8K this shift is equal to 1.5meV whereas at
temperatures above $T$=3.6K it is about 1.1meV only. The PL line
width evolution is highly sensitive to temperature: At 3.8K such a
strong narrowing of the line with increasing delay does not occur
as it is observed T = 1.8K.

   The two decay curves shown on figure 3 correspond to different
excitation powers, measured at the PL line maximum of the IEs at
$T$=2K and $U$=0.6V. One can see that decay time for larger power
is much shorter (about 3 versus 7ns) than for smaller one. This
description is approximative and has been used for the
monoexponential data fitting.

\begin{figure}[h]
\begin{center}
\includegraphics[width=8.5cm]{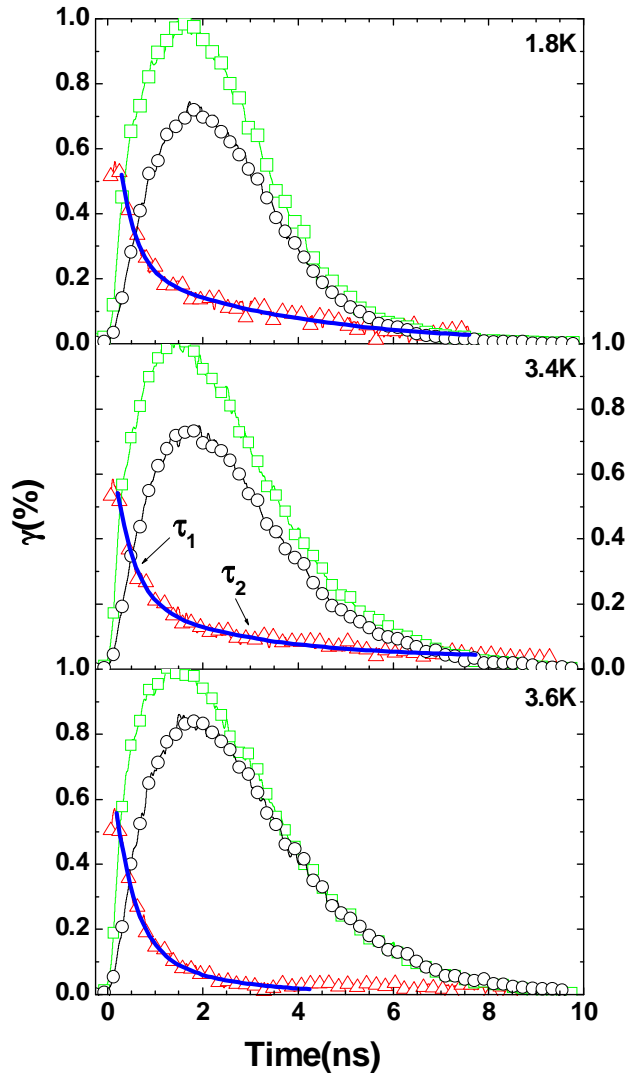}
\end{center}
\caption{Circular polarization resolved decay curves of the IEs PL
($\sigma^{+}$ - green curves, $\sigma^{-}$ - black curves) and
circular polarization degree of IEs PL (red curves) measured at
different temperatures within the crosshatched region in Fig. 2.
$U$ = 0.6 V.}
\end{figure}

The time evolution of the circular polarization degree reflects
the IE spin relaxation. The circular polarization is defined as
$\gamma = (I_{\sigma^{+}} - I_{\sigma^{-}})/(I_{\sigma^{+}} +
I_{\sigma^{-}})$, where $I_{\sigma^{+}}, _{\sigma^{-}}$ are the PL
signal intensities for the $\sigma^{+}$ and $\sigma^{-}$
components. From Fig. 2 it becomes evident that $\gamma$ varies
with the PL emission energy and also depends strongly on
temperature. Within the first nanosecond after the laser pulse,
the IE PL line is strongly circularly polarized (more than 60 \%
at zero delay, for example) and does not show strong variations
across the whole PL spectrum. At later times only the high-energy
part of the spectrum remains polarized. At low temperature T=1.8K
the circular polarization of the PL disappears after 7ns while at
T=3.8K the PL is unpolarized already after 3ns.

For a quantitative analysis of the spin relaxation, the PL image
is profiled along the energy scale, from which PL decay curves for
a fixed energy interval are received. The spin relaxation time is
the decay time of the circular polarization degree. Figure 4 gives
the IEs PL decay curves for $\sigma^{+}$ and $\sigma^{-}$ circular
polarization (circular and square symbols, respectively) as well
as the decay of the corresponding circular polarization degree
(triangular symbols), measured at different temperatures near the
PL line maximum (crosshatched region in fig.2). The blue lines are
least-mean-square fits to the data using a bi-exponential decay
form.

The maximum intensity of the IE PL line is reached for delays of
about 2ns. We suggest that this time is necessary for the
formation of IEs upon resonant tunneling of electrons and holes to
adjacent quantum wells and relaxation in energy towards the
density and temperature equilibrium values.

We have found that the spin relaxation dynamics can be described
by two different time constants, an initial fast one and a delayed
slow one (see fig.5). The initial decay time $\tau_{1}$ is very
weakly temperature dependent and amounts to about 0.35ns. In
contrast, the slow decay time $\tau_{2}$ which exceeds $\tau_{1}$
by an order of magnitude at low $T$ drops by a factor of about 2
for temperatures above 3.6K. Up to 15K no more considerable
changes in the temporal dynamics of the IEs circular polarization
degree is observed. The same behavior occurs for another, slightly
smaller bias $U$ = -0.55 V applied to the CQW. These data have
been received at an excitation power about 30kW/cm$^{2}$ leading
to an IEs concentration $n_{ex}\sim 3 \times 10^{10}$cm$^{-2}$. We
have found that the strong reduction of spin relaxation time
$\tau_{2}$ is quite sensitive to excitation power (see fig.6). At
smaller and at bigger excitation power the temperature boundary
for the strong reduction of the spin relaxation time $\tau_{2}$
shifts to lower temperatures.

Fig.7 shows the bias dependence of the spin relaxation time at
T=2K. With increasing voltage, $\tau_{1}$ and $\tau_{2}$ first
monotonously increase and then don't change up to 0.85V,
corresponding to about 22meV splitting between the 1sHH and the
interwell exciton. For higher bias, the spin relaxation process is
described by a single relaxation time. From Fig.6 it can be seen
that with increasing bias the time for accumulation of interwell
excitons increases. This can be attributed only due to a rise of
the tunneling time. Therefore we suggest that for large bias
single-particle spin-relaxation mechanisms may play a
determinative role.

We have also measured the temperature dependence of the circular
polarization degree of the intrawell 1sHH exciton luminescence.
Its spin relaxation rate does not change in the temperature
interval from 2 to 15K and decays monoexponentially with a time
constant of about 180ps. This result is in  good agreement with
ref. [7], where the exciton spin relaxation dynamics has been
investigated in great detail at low temperatures. The mechanism
responsible for 1sHH spin relaxation is the electron-hole exchange
interaction. For CQW structures the carries spin dynamics is more
complicated. After the laser pulse several processes may occur:
electron tunneling to the adjacent quantum well, energy and spin
relaxation, interwell exciton formation and radiative
annihilation. We suggest that the $\tau_{1}$ time is due to
electron-hole exchange interaction within the 1sHH exciton, while
the $\tau_{2}$ time characterizes the IEs spin relaxation. Since
the wave function overlap integral is very small, the
electron-hole exchange interaction is very weak and the $\tau_{2}$
time is much longer than the $\tau_{1}$ time.

\section{Theoretical description of interwell exciton spin relaxation.}

The exciton kinetics including spin relaxation is governed by the
following processes:

1) electrons spin-flip within the exciton with rate $w_e$,

2) holes spin-flip within the exciton with rate $w_h$,

3) exciton spin-flip due to electron-hole exchange with rate
$w_{EX}$ for intrawell excitons and with probability $w_{ex}$ for
IEs,

4) exciton radiative recombination with rate $w_R$ for intrawell
excitons and with probability $w_r$ for IEs,

5) intrawell exciton transformation into IEs due to electron
tunneling to the adjacent quantum well with rate $w_k$.

For modelling, we have used the equations obtained in refs. [10,
11]. The concentrations of intrawell excitons $N^D_i$  and
interwell excitons $N^I_i$ are described by the rate equations:
\begin{eqnarray}
\frac{d}{dt}N_i^{D}=F_{ij}^{D}N_j^D-w_{k}N_i^{D}, \\
\frac{d}{dt}N_i^{I}=F_{ij}^{I}N_j^I+w_{k}N_i^{D},
\end{eqnarray}
where $i,j=3/2, 1/2, -1/2, -3/2$

These equations have to be solved for the boundary conditions:
\begin{eqnarray}
N_1^D=N_0,~~N_i^{D}=0,~i\neq 1,~~N_i^{I}=0. \nonumber
\end{eqnarray}
The coefficients $F_{ij}^{D/I}$ are given by
\begin{equation}
F^{D(I)}_{ij}=
 \left|
\begin{array}{llrr} -(w_e^{+}+w_h^{+})& w_e^{-}& w_h^{-}&0\\
w_e^{+}&W^{D(I)}&w_{EX(ex)}&w_h^{+}\\w_h^{+}&w_{EX(ex)}&W^{D(I)}
&w_e^{+}\\0&w_h^{-}&w_e^{-}&-(w_e^{+}+w_h^{+})
\end{array}
\right|.
\end{equation}
The intra- and interwell exciton spin flip probabilities are
\begin{equation}
W^{D(I)}=-(\frac{1}{\tau_{R(r)}}+w_{EX(ex)}+w_e^{-}+w_h^{-}).
\end{equation}
The electron (hole) spin-flip rate inside the exciton is given by
\begin{equation}
w_{e(h)}^{\pm}=-\frac{w_{e(h)}}{1+e^{\pm \Delta/kT}},
\end{equation}
where $\Delta > 0$ is the energy splitting between the optically
active and the optically inactive ones.

\begin{figure}[h]
\begin{center}
\includegraphics[width=8.5cm]{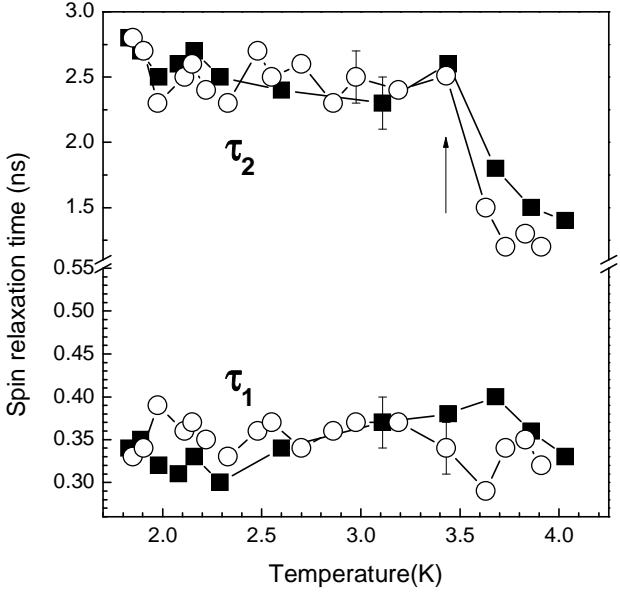}
\end{center}
\caption{Temperature dependence of spin relaxation times. The
circular symbols give the fast $\tau_{1}$ and the slow $\tau_{2}$
relaxation times at bias $U$=0.6V, the square symbols give the
$\tau_{1}$ and the $\tau_{2}$ times at $U$=0.55V, respectively.
The arrow indicates the temperature region where a rapid change of
the spin relaxation rate occurs.}
\end{figure}

We assume that the hole spin flipping time is much smaller than
all other times in our system. In this case the "bright" ($|\pm
1>$) and "dark" ($|\mp 2>$) exciton concentrations are connected
to each other, depending on temperature $T$ and exchange splitting
$\Delta$.
\begin{equation}
N^{\pm 2}=N^{\mp 1}f(\Delta /kT),~~f=\frac{1+e^{\Delta
/kT}}{1+e^{-\Delta /kT}}.
\end{equation}
This assumption allows one to take into consideration only the
optical active ("bright") $|\pm 1>$) excitons.
\begin{eqnarray}
\frac{d}{dt}N_i^{D}=G_{ij}^{D}N_j^D,
\end{eqnarray}
with
\begin{eqnarray}
N_1^D=N_0,~~N_i^{D}=0,~i=\pm 1 \nonumber
\end{eqnarray}
\begin{eqnarray}
\frac{d}{dt}N_i^{I}=G_{ij}^{I}N_j^I+w_{k}N_i^{D},
\end{eqnarray}
with $N_i^{I}=0$ and the coefficients $G^{D(I)}_{ij}$ given by
\begin{equation}
G^{D(I)}_{ij}=
 \left|
\begin{array}{lr} -(w_{L(l)}+w_{X(x)})& w_{X(x)}\\
w_{X(x)}&-(w_{L(l)}+w_{X(x)}).
\end{array}
\right|
\end{equation}
The influence of the "dark" $|\pm 2>$ excitons is described by the
effective annihilation rates $w_L$, $w_l$ and the spin relaxation
rates $w_X$, $w_x$ (effective times $\tau_L$, $\tau_l$ and
$\tau_X$, $\tau_x$) for intrawell and IEs, respectively:
\begin{eqnarray}
\nonumber w_L&=&w_k+w_R/(1+f), \\ \nonumber w_l&=&w_r/(1+f), \\
\nonumber
w_X&=&2w_{EX}/(1+f)+4w_e/(2+e^{\Delta /kT}+e^{-\Delta /kT}), \\
w_x&=&2w_{ex}/(1+f)+4w_e/(2+e^{\Delta /kT}+e^{-\Delta /kT}).
\end{eqnarray}

To compare these results with the experimental data it is
necessary to determine the time dependence of the exciton
concentration of the optically active states $|\pm 1>$ and of
their spin polarization degree:
\begin{eqnarray}
p(t)&=&\frac{N_{+1}(t)-N_{-1}(t)}{N_{+1}(t)+N_{-1}(t)} \\ N_{\pm
1}(t)&=&\frac{1}{2}N(t)(1 \pm p(t)),
\end{eqnarray}
where $N(t)=N_{+1}(t)+N_{-1}(t)$ is the total exciton
concentration. The IE PL intensities measured in experiment are
directly proportional to the exciton concentrations. For intrawell
excitons we obtain:
\begin{equation}
p^{D}(t)=e^{-w_{X}t};~~~N^{D}(t)=N_0e^{-w_{L}t},
\end{equation}
while for IEs we have:
\begin{eqnarray}
p^{I}(t)&=&\frac{w_L-w_l}{w_L+w_{X}-w_{l}-w_{x}}\frac{e^{-(w_{l}+w_{x})t}-e^{-(w_{L}+w_{X})t}}{e^{-w_{l}t}-e^{-w_{L}t}}
\nonumber
\\ N^{I}(t)&=&\frac{N_0w_k}{w_L-w_l}(e^{-w_{l}t}-e^{-w_{L}t}).
\end{eqnarray}
The comparison of experimental data with theoretical calculations
is presented in Fig.8 for different electrical biases
corresponding to the following energy splittings $\Delta E$
between the 1sHH exciton and the interwell exciton: (1) $\Delta
E=4.5$ meV, $\tau_L= 0.16$  ns, $\tau_l=0.8$ ns, $\tau_X=0.12$ ns,
$\tau_x=1.1$ ns; (2) $\Delta E=6.4$ meV, $\tau_L=0.20$ ns,
$\tau_l=1.0$ ns, $\tau_X=0.13$ ns, $\tau_x=1.6$ ns, and (3)
$\Delta E=8.3$ meV, $\tau_L=0.27$ ns, $\tau_l=1.3$ ns,
$\tau_X=0.16$ ns, $\tau_x=2.0$ ns. Our model rather well describes
spin relaxation at small bias, while at bigger voltages probably
it would have to take into account non-local tunneling, to obtain
better agreement.

The increase of the exciton spin relaxation time and the radiative
exciton annihilation time with bias can be explained by the
enhanced spatial separation of electrons and holes both in the
same quantum well as well as in adjacent ones.

\begin{figure}[h]
\begin{center}
\includegraphics[width=8.5cm]{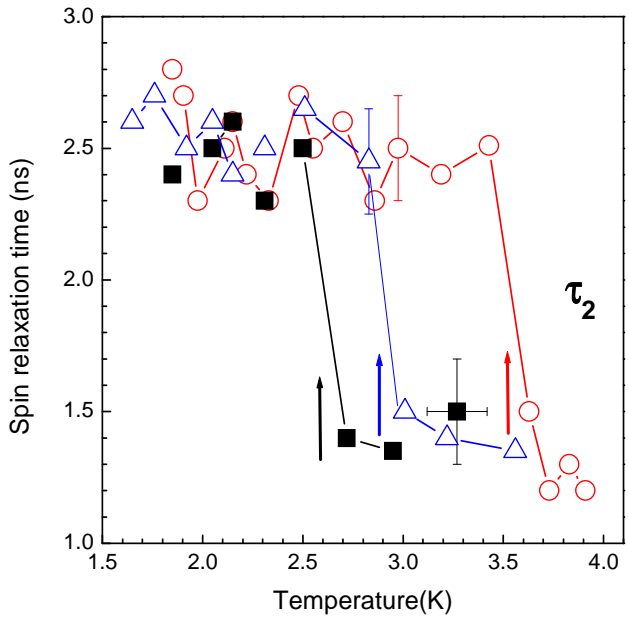}
\end{center}
\caption{Temperature dependence of the spin relaxation time
$\tau_{2}$ at different excitation powers. Red circular symbols
correspond to an excitation power of about 30kW/cm$^{2}$ leading
to an IEs concentration $n_{ex}\sim 3 \times 10^{10}$cm$^{-2}$.
Black and blue square symbols correspond to excitation powers of
about 20kW/cm$^{2}$ and 45kW/cm$^{2}$, respectively. The arrows
indicate the temperature region where a rapid change of the spin
relaxation rate occurs.}
\end{figure}

\section{Discussion}

We believe that the experimental results described above strongly
support the suggestion that we have made previously (see [7])
about the collective nature of the behavior of the interwell
excitons below a critical temperature. Qualitatively, the origin
of the collective exciton phase can be described as follows. At
low temperatures (T $\leq$ 2K), the IEs fill a particular
potential relief in the quantum-well plane, as the density of the
optical excitation power is increased. These potential traps
arise, for example, from residual impurities, defects, or other
structural imperfections. This is manifested by a narrowing of the
PL line with increasing pumping, such that the line width ceases
to reflect the statistical distribution of the fluctuation
amplitudes of the random potential. In our opinion, the sharp
narrowing of the PL line, the superlinear rise of its intensity
and the threshold-like increase of its circular polarization
cannot be associated with reaching the percolation threshold by IE
density only, because of the strong sensitivity to temperature,
even though there is no distinct temperature boundary. Berman and
Lozovik showed [2] that a sufficiently dense system of IEs with
particular values of the dipole moment may condense into a
dielectric phase despite of the dipole repulsion among such
excitons.

\begin{figure}[h]
\begin{center}
\includegraphics[width=8.5cm]{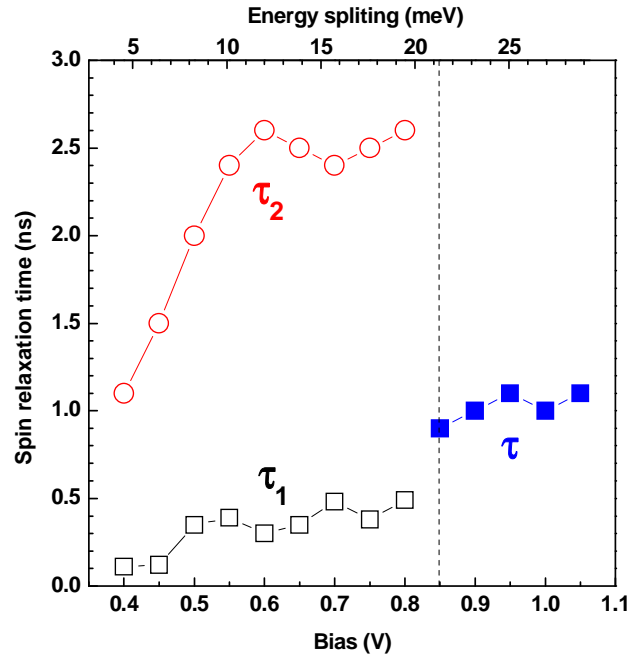}
\end{center}
\caption{Bias dependencies of the spin relaxation times at $T$=2K.
Red circular symbols correspond to the slow $\tau_{2}$ relaxation
times, black square symbols to the fast $\tau_{1}$. The blue
square symbols give the spin relaxation times $\tau$ for high bias
where a monoexponential decay of the circular polarization degree
is observed. The dashed line indicates the boundary between the
biexponential and the monoexponential spin relaxation ranges.}
\end{figure}

An essential amendment was made in ref. [1], whose authors
indicated that such a condensation in real systems can occur most
probably in regions with confinement in the quantum-well plane. In
the structures studied here which were fabricated using the growth
interruption epitaxial technique for the heteroboundaries (in our
case, the growth interruption time was 2 minutes), large-scale
in-plane fluctuations of the well width arise in the
heterostructure plane.  The size of these fluctuations along the
growth direction is of the order of one monolayer. The
characteristic lateral length scale of such fluctuations in the QW
plane reaches a micrometer. Because of these fluctuations, lateral
domains are formed in the quantum wells. As judged from the
characteristic doublet structure in the photoluminescence
excitation spectra of the intrawell excitons of our samples, the
depth of such domains can be estimated by 1.5 - 2 meV. IEs can
accumulate in these domains, because the lateral domain boundaries
prevent excitons from spreading out randomly in the quantum-well
plane.

We have suggested that the IEs demonstrate a collective behavior
in these domains when their density and temperature surpass
critical values. For testing this assumption, the surfaces of the
samples were coated by a metallic mask containing lithographically
prepared holes with sizes of a micrometer or less, through which
photoexcitation and photocollection were done [8] (see fig. 9). We
found that for weak pumping (less than $50\mu W$), the IEs are
strongly localized in small-scale fluctuations of a random
potential, and the corresponding photoluminescence line is
inhomogeneously broadened (up to 2.5meV). When the resonant
excitation power is increased, a spectral line which is attributed
to delocalized excitons arises with a threshold-like intensity
behavior. Above the threshold the intensity increases linearly
with pump power, narrows (minimum line width $350\mu$ eV), and
undergoes an energy shift (up to $\sim 0.5$meV) to lower energies,
in accordance with filling of the lowest state in the domain.
Finally we observed that with increasing temperature, this line
disappears continuously from the spectrum. Therefore its vanishing
cannot be described by a thermally activated behavior, which would
show an exponential dependence. The critical temperature $T_{c}$
is about 3.4K.

\begin{figure}[h]
\begin{center}
\includegraphics[width=8.5cm]{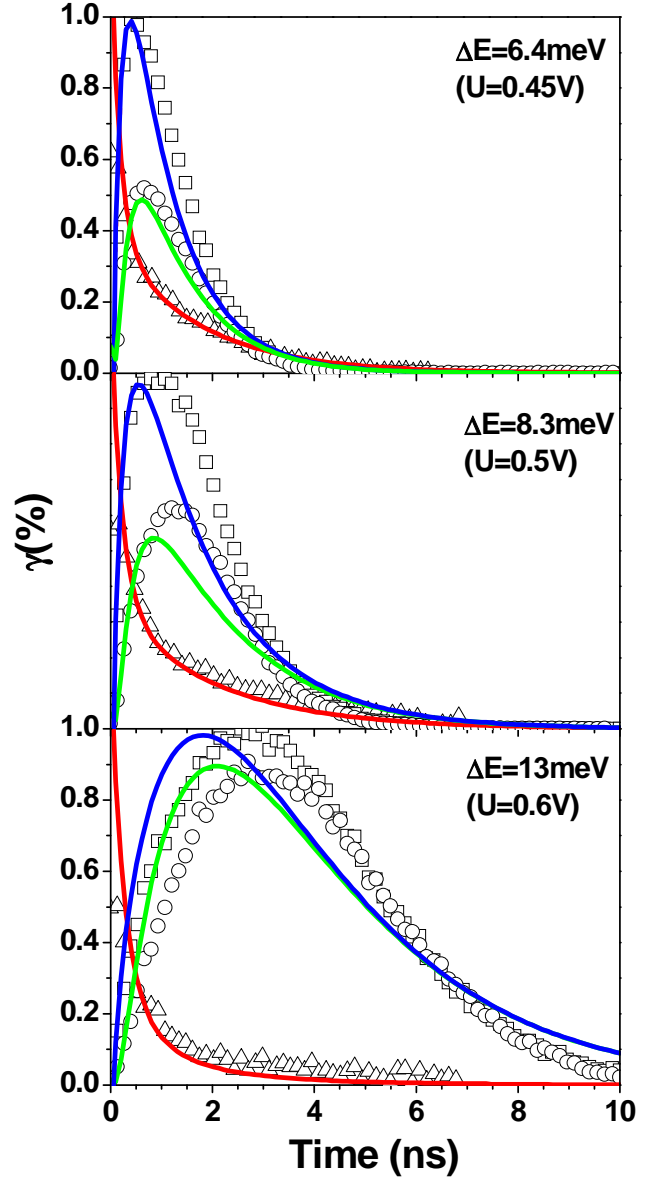}
\end{center}
\caption{Circularly polarized IEs PL decay curves ($\sigma^{+}$ -
square symbols, $\sigma^{-}$ - circular symbols) as well as the
corresponding circular polarization degree $\gamma$ (triangular
symbols) measured at different biases. The signal has been
integrated over the crosshatched region in fig.2. Solid curves
correspond to theoretical calculations according to expressions
(14)-(15). T=2K.}
\end{figure}

IEs are composite bosons. If this bosonic behavior is maintained
also at high densities, the excitons therefore must condense upon
reaching the values for critical concentration and temperature
(analogue of Bose-Einstein condensation). For confinement in the
quantum-well plane, the critical temperature at which this
condensation takes place can be estimated using the equation
$T_{c} = \pi \hbar^{2} N_{ex}/km_{ex}\ln(N_{ex}S)$, where $N_{ex}$
is the exciton density, $m_{ex}$ is the exciton mass, and $S$ is
the domain area. If we assume that the exciton mass $m_{ex}$ is
$0.25m_{0}$ and the domain size is $0.5\mu^{2}$, we obtain a
critical temperature $T_{c} = 3K$ for the densities $N_{ex}=
5\times 10^{10} cm^{-2}$ used in our experiment. This is very
close to the value observed experimentally.

It should also be noted that in our experiments measurements are
carried out simultaneously on several tens of such lateral domains
because the smallest diameter of the laser excitation spot from
which luminescence spectra are detected is about $30\mu m$. Due to
the dispersion of the lateral domain sizes and the averaging of
spectra from different domains, a sharp threshold of the critical
behavior in temperature cannot be observed. For the same reasons,
the smallest observed luminescence line width (about 1meV) is
still inhomogeneously broadened, because interwell exciton
luminescence from differently sized domains contribute to the
emission. At the same time, the sharp narrowing of the
luminescence line observed experimentally at $T < T_{c}$
($T_{c}=4K$) and the long-wavelength shift of this line (about
1.5meV), in accordance with filling of the lowest energy state in
the domain, are clear manifestations of Bose properties of
excitons.

\begin{figure}[h]
\begin{center}
\includegraphics[width=8.5cm]{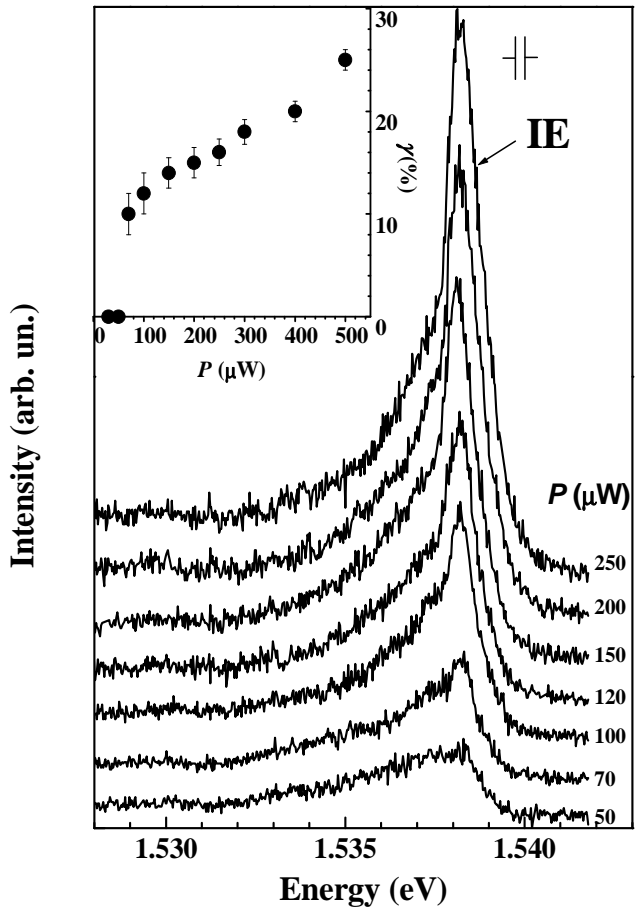}
\end{center}
\caption{Photoluminescence spectra of the interwell exciton ($IE$
line) under conditions of resonant excitation of the direct 1sHH
exciton for various excitation powers. $U$=0.3V, $T$=1.6K. The
used excitation power is indicated at each trace. The spectral
resolution of the setup is indicated at the top right. Inset:
degree $\gamma$ of circular polarization (dots) vs. excitation
power for resonant excitation at the line maximum; the
contribution from the structureless background was not taken into
account. Again, $T$=1.6K and $U$=0.3V.}
\end{figure}

The condensed IE phase must exhibit coherent properties. This
means that the IEs must possess the same phase on the length scale
of the de Broglie wavelength, which is close to the lateral domain
sizes. This phase coherence in turn affects the radiative
annihilation rate, which increases due to the increased coherence
volume. Earlier we have found from the kinetics of luminescence
spectra that the lifetime of the collective exciton state is about
three times shorter than the luminescence decay time of localized
IE's [9]. This increase in the radiative decay rate of IEs and the
corresponding increase of the degree of circular polarization are
particular manifestations of the coherence of the collective
exciton state.

We believe that the presented experimental results are additional
evidence for the coherence of the IEs collective phase at low
temperatures. Our claim is based on ref. [12], where the spin
relaxation rate of Bose-condensates of atoms in traps was studied.
It has been shown that the spin relaxation rate of the atoms
condensed phase is N! times smaller than that for atoms in the
uncondensed phase, where $N$ is the number of particles involved
in scattering process destroying Bose-condensation.
Experimentally, this claim was confirmed in ref. [13], in which
the spin dynamics of atoms in a Bose-condensate has been
investigated. In our case the $\tau_{2}$ time characterizing the
exciton spin relaxation changes by a factor of about 2, as
expected from the electron-hole composition of the exciton, in
good agreement with the model calculations in ref. [12].

\section{Conclusions}

In conclusion, for the first time the temperature dependence of
the IEs spin relaxation time has been investigated in GaAs/AlGaAs
CQWs. A strong decrease of the IEs spin relaxation time has been
discovered at $T < T_{c}=4K$. The observed phenomenon occurs due
to the interwell exciton collective phase coherence at
temperatures below the critical one.

\section{Acknowledgements}

The authors want to express their thanks to V.B.~Timofeev for
valuable suggestions and remarks, as well as to Yu.~Kagan for
fruitful and interesting discussions. This research was supported
by DFG (Grant No. 436 RUS 17/95/03), by RFBR, and by the Russian
Science Support Foundation.

\end{document}